\documentclass[journal]{IEEEtran}
\usepackage{graphicx}
\usepackage{graphics}
\usepackage{caption}
\usepackage{algorithm}
\usepackage{algpseudocode}
\usepackage{balance}
\usepackage{amsmath}
\usepackage{bm}
\usepackage{url}
\usepackage{amssymb}
\usepackage{multirow}
\usepackage[table]{xcolor}
\usepackage{color}
\usepackage{cite}
\usepackage{caption}
\usepackage{subfigure}

\begin{document}

\title{The Paradigm of Digital Twin Communications}
\author{Tom H. Luan, Ruhan Liu, Longxiang Gao, Rui Li, Haibo Zhou}
\maketitle

\begin{abstract}
With the fast evolving of cloud computing and artificial intelligence (AI), the concept of digital twin (DT) has recently been proposed and finds broad applications in industrial Internet, IoT, smart city, \emph{etc}. The DT builds a mirror integrated multi-physics of the physical system in the digital space. By doing so, the DT can utilize the rich computing power and AI at the cloud to operate on the mirror physical system, and accordingly provides feedbacks to help the real-world physical system in their practical task completion. The existing literature mainly consider DT as a simulation/emulation approach, whereas the communication framework for DT has not been clearly defined and discussed. In this article, we describe the basic DT communication models and present the open research issues. By combining wireless communications, artificial intelligence (AI) and cloud computing, we show that the DT communication provides a novel framework for futuristic mobile agent systems.
\end{abstract}


\IEEEpeerreviewmaketitle

\section{Introduction}

\label{section:intro}

\begin{figure*}[tbp]
\centering
\includegraphics[width=\linewidth]{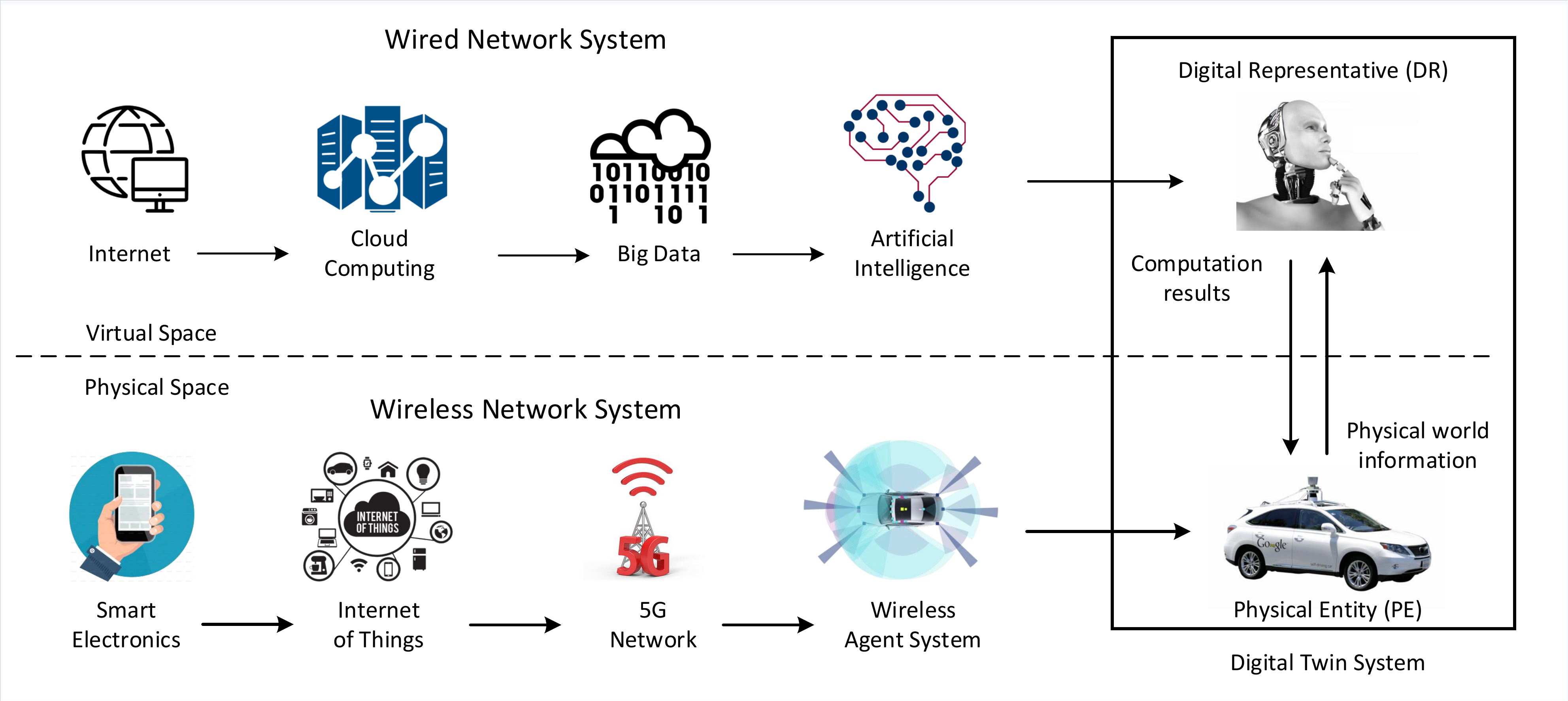}
\caption{Technology lines of networking}
\label{fig: two_line}
\end{figure*}

Over the past decades, the networking technology has evolved along two
lines, as illustrated in Fig.~\ref{fig: two_line}.

The first line is the development of the wired network technology from the
Internet to cloud computing, big data analysis and AI. Notably, with the
continuous evolution of Internet and wired access technologies starting from
1990s, as of April 2021, the population of Internet users reaches 4.72 billion, which is 57 percent of the world population. The intensive computing demands of Internet users drive the development of the centralized cloud computing and data center networks. As reported, by 2021, 94 percent of workloads and compute instances are processed by cloud data centers, making the cloud as the centre of data and computing in the worldwide. Finally, the global connectivity, giant data resource
and rich computing power, coupled with the advances of artificial
intelligence (AI), make the cloud a supreme platform nowadays for the global intelligence.

Another line is the surge development of the wireless network technologies
from the smart electronics to Internet of Things (IoT), fifth-generation
(5G) cellular network and mobile agent systems. The wide adoption of smart
electronics, \emph{e.g.}, iPhone, represents the first step to put the
computing power on the palm of people. As a result, the way that people live has
been revolutionized with various mobile applications and the IoT technology,
such as smart home and smart community \cite{li2011smart}. With the explosive usage of
wireless electronics, 5G networks have emerged to provide the larger network
capacity, smaller communication latency and higher spectrum efficiency to
support the massive ubiquitous mobile devices. The fast and ubiquitous
wireless connectivity, powerful onboard computing facility and AI have paved
the way for mobile agent systems, \emph{e.g.}, UAVs, autonomous vehicles,
which open the potentials for a high-degree community-wide automation.

It is foreseeable that the broad applications of mobile agent systems would
result in another revolution in our lives towards the full automation and
significant enhancement on the life quality, efficiency and convenience.
However, the mobile agents are constantly limited by their battery, onboard
processing power and sensing range. To alleviate the physical limitations of
mobile devices, the line of wireless networking technologies needs to seek
for the rich AI and resources from the cloud, which drives the two lines
converge at the DT technology. Specifically, a DT system maintains a mirroring digital
representative (DR) at the virtual space, \emph{e.g.}, the cloud for a mobile agent (referred to as the physical entity
(PE)). In this framework, the
PE works in the real-world environment for practical task executions. The DR
keeps real-time communications with their PE to learn PE's status. With such
information, the DR works on the cloud as a collaborating peer to the PE to
exploit the cloud resource to help the PE on finishing its tasks.

The concept of DT has been developed for decades. Nowadays, DT has found
rich applications in industry with the smart grid system, Industry 4.0,
smart city and healthcare. However, in contrast to most existing works which
consider DT as a simulation platform, a thorough investigation on DT as a
promising communication framework to facilitate the mobile agent
systems is rare in the literature. In this article, we introduce the basic
concepts and characteristics of DT communication, discuss the possibility of
different application scenarios. For further study, we explore the three
ways of DR deployment and analyze how it differs from other technologies.

\section{Structure of DT and its Communication Model}

A typical DT concept model contains four parts: a PE in real space, a
DR in virtual space, the connections of current data and information that
ties the PE and DR together, namely intra-twin
communications, and the connections to connect the DR with the outside world. The functions and characteristics of each
part is described as follow.

\paragraph{Physical Entity (PE)}

In the real space of a DT system, various of infrastructures such as sensors
and cameras are responsible for collecting data of current physical
measurements of PE. Taking the autonomous vehicles as
an example, as in Fig.~\ref{fig: two_line}, the PE is a real-world vehicle
containing different sensors, \emph{e.g.}, the cameras which are installed at every angle
and ensures the 360 degree view of the external environment, and radar and LiDAR on the four sides of the vehicles can be used to detect the objects around and measure its
distance and speed in real time.

\paragraph{Digital Representative (DR)}

In the virtual space of a DT system, a DR can be treated as a software
application maintaining a real-time model of its PE. It receives the
real-time data collected by PE via a synchronized private link. After
processing the data, the DR can not only visualize and represent the instant
status of its PE, but also can calculate anticipatory operations to help its
PE to make decisions in the real-world. For example, as in Fig.~\ref%
{fig: two_line}, according to the information received from a vehicle, \emph{%
e.g.}, the traffic environment, weather conditions, passenger's preferences,
and the information learned from the cloud such as regional traffic
information, a DR of a vehicle can calculate the best driving strategy on
the velocity, lane to stay and feedback to the vehicle. The information from
the DR help the vehicle to make the most accurate path planning and optimal
driving strategy.

\paragraph{Intra-twin Communications}

The connection between PE and DR, \emph{i.e.}, the intra-twin communication, contains two components: the raw data transmission and the processed information transmission.
Specifically, the raw data refers to the data collected by different sensors
that flow from PE to the DR, whereas the processed information refers to
the analysis result that generated and sent from DR to the PE. What is
noteworthy is that the connection link between PE and its DR has to be
private and completely protected to share the information and knowledge (or
AI models) within the twin. The three components, \emph{i.e.}, PE, DR and intra-twin communication, make up a complete inner loop of the DT system that can help implement effective simulation, prediction and feedback loops within the twin.

\paragraph{Communications to Outside World}

The connection of DRs to the outside world include two types.

\begin{figure}[tbp]
\centering
\includegraphics[width=1\linewidth]{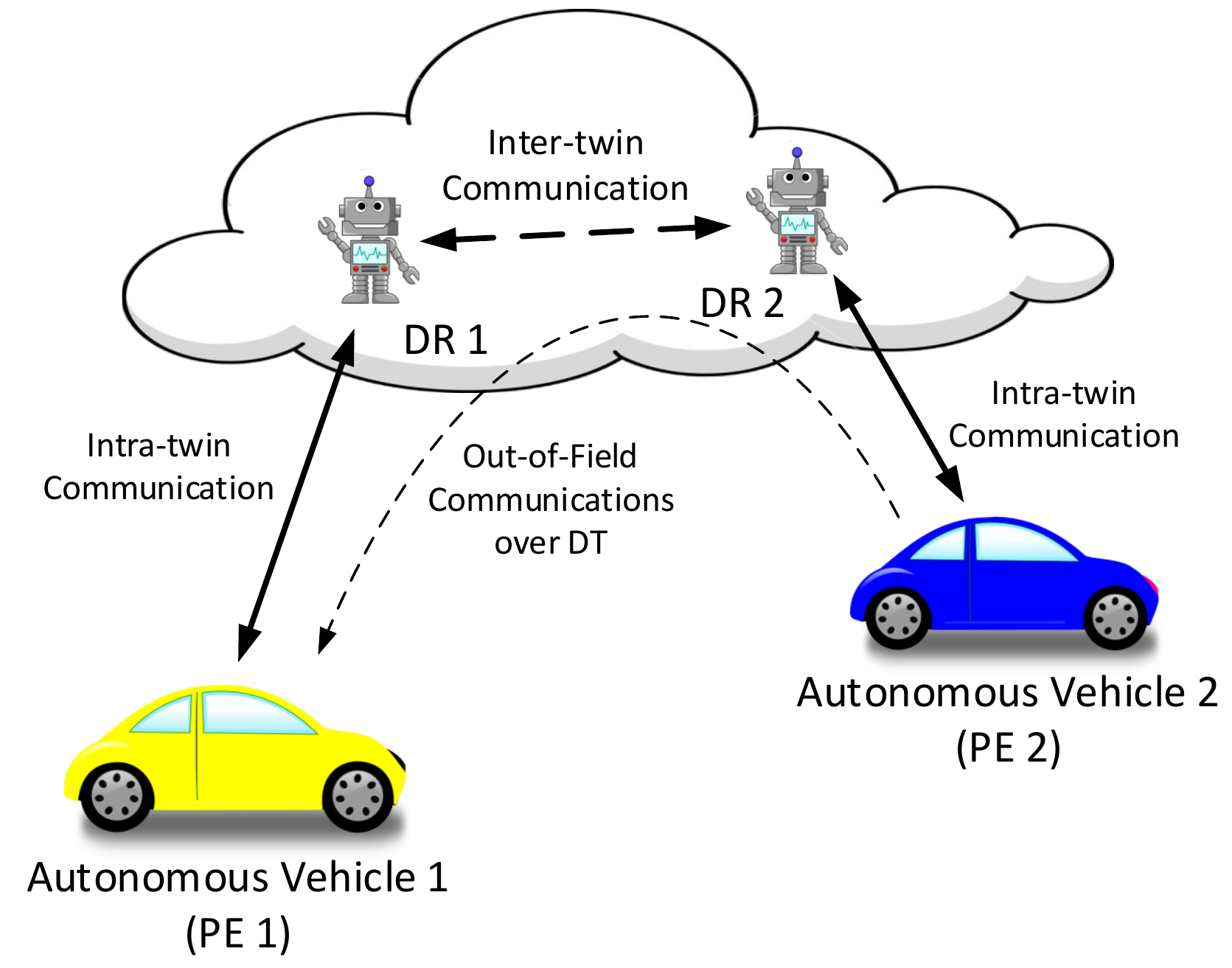}
\caption{DT system for autonomous vehicles}
\label{fig: av_dt}
\end{figure}

The first type is the communications among DRs, \emph{i.e.}, inter-twin
communications. Note that with each DR keeping the synchronized communications
with its PE, the communications among PEs can be relayed through their DRs. For
example, as in Fig.~\ref{fig: av_dt}, assuming that PE~1 intends to learn
the road information from PE~2 in its driving path. The direct
vehicle-to-vehicle communications between PE~1 and PE~2 may be unavailable due to
the out-of-field. To achieve this goal, DR1 can communicate with DR~2
to retrieve PE~2's sensed information using the inter-twin communication, and transmit to PE~1 using its intra-twin communication.

The second type is the communications of DR with the cloud. By living on the
cloud, the DR can retrieve information from cloud and synchronize with its
PE. In the meantime, the cloud can also communicate with DRs to retrieve
their PE's information for census.

\section{Characteristic of DR}

As a key component of DT, the DR is a new digital approach that communicate,
compute, simulate and control the whole life cycle of PEs by using PE's
historical data, real-time data, cloud data and algorithmic AI models \cite%
{Taofei2018}. According to this concept, the typical characteristics of a DR
can be described as follows:

\begin{itemize}
\item \emph{Always-on Application}: A DR can utilize the computing and
storage resources at the cloud to assist its PE. A PE can be off-line due to
limitations on wireless connectivity. In this case, a DR can work on behave
of its PE in the cloud.

\item \emph{Synchronization}: In contrast to simply a cloud computing scheme, a
DR keeps the real-time bidirectional information synchronization with its
PE. On one hand, the PE updates the real-time local information collected at
the physical end. At the same time, the DR may also collect useful
information from other DRs or the cloud, and synchronize with the PE to help PE on its
task accomplishment.

\item \emph{Private}: The goal of a DR is to assist its PE, and therefore,
DR works as a collaborating peer to its PE. Therefore, as a synchronized
pair, a DR is private to the PE and cannot release any sensitive information
of its PE.

\item \emph{Autonomous}: With sufficient data and computing power, a DR can
work autonomously. For example, in the inter-twin communications in Fig.~\ref{fig: av_dt}, DR~1 can communicate with DR~2 without notifying PE~1, and only
synchronize the retrieved information to PE~1. The autonomous behavior of
DRs should acquire the permissions of their PEs.
\end{itemize}

%
%
%
%

\section{State-of-the-Art}

With the development of networking and AI technologies, the state-of-the-art research frames DT as a modern imperative for digital transformation initiatives. Three typical application scenarios of DT system in literature is introduced as follows.

\subsection{Model of Communications}

As a new paradigm that mainly used in the cyber-physical environment, how to ensure high quality communication capability is the basic topic of a DT system.
As illustrated in Fig.~\ref{fig:scenario}, the DT systems focusing on the digital operations in the virtual space and the existing network technologies in the real-world, such as 5G and fiber grid, can be complementary to each other. Specifically, the impact between the two is mutual, by establishing DR of target network, the performance of the network in terms of security, data communication rate, and bandwidth can be greatly improved. Meanwhile, with the aid of high-performance network, DR can meet the requirements for synchronization and privacy in simulation and computing scenarios.

At present, there are several studies exploring how the characteristics of DT can be used to improve the communication quality of network. In order to solve the mismatch between the conventional network management scheme and the real physical situation, Wang \textit{et al.} \cite{9356524} propose a DT based optical communication system to narrow the gap between the physical layer and network layer, so as to achieve a high-reliable and high-efficiency optical communication system. To accommodate the development of smart vehicles, a DT based social-aware vehicular edge network is introduced by \cite{9399641} to conduct massive content delivery. By constructing a social model in DR, a cache cloud is generated to merge corresponding stored content from multiple vehicles, which shows great advantages in optimizing caching performance. In industrial IoT environment, Jia \textit{et al.} \cite{9214874} propose an intelligent clock synchronization method. With the predictable clock behaviors, the synchronization-related timestamp exchange is significantly reduced, and a high quality data exchange network with high clock accuracy and low communication resource consumption can be derived.



\subsection{Simulation and Prediction}



\begin{figure*}[tbp]
\centering
\includegraphics[width=.8\linewidth]{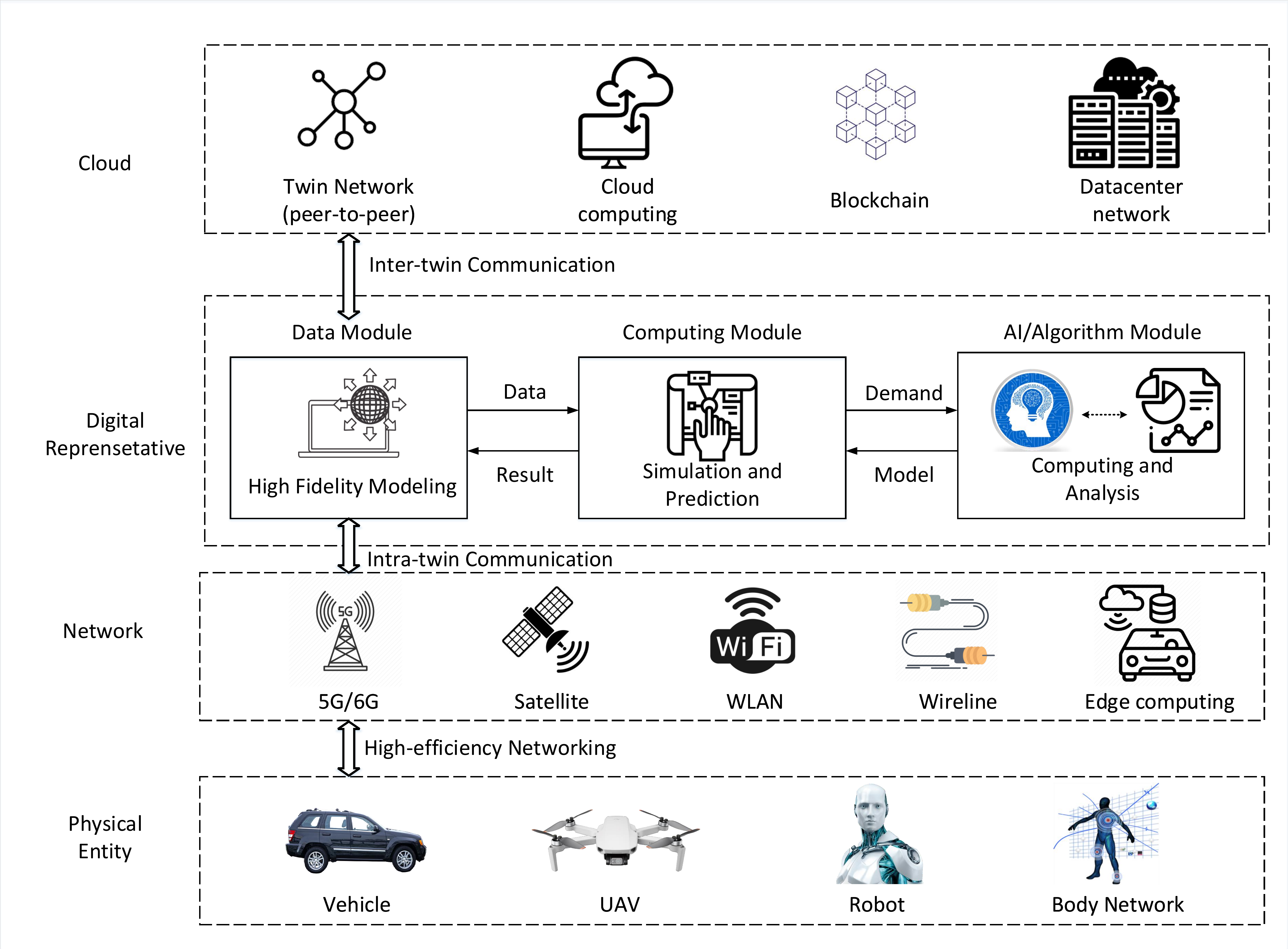}
\caption{Application scenarios of DT communication system}
\label{fig:scenario}
\end{figure*}

With high-quality communication as the basis to transmit data and knowledge, the DT system can be used to perform simulation tasks in a variety of scenarios, such as anomaly detection and autonomous vehicles testing. This is particularly useful for areas that involve a mass of labor or resource consumption. As shown in Fig.~\ref{fig:scenario}, from a holistic perspective, a simulation can be seen as a bridge in the whole DT ecosystem, that is, it is directly related to the underlying application scenario of DT, \emph{i.e.}, communication, to ensure accurate and real-time data transmission. At the same time, simulation also maintains massive information exchange with the upper application scenario, \emph{i.e.}, computing and analysis, in order to fulfill the real-time task requirements in the real-world environment.

Benefit from the fewer constraints, simulation is the closest scenario to industrial-grade applications and has received a lot of attention from enterprises and research institutes. Danilczyk \textit{et al.} \cite{9000371} develop a DT based security framework named ANGEL to intelligently guard the physical system of the smart grid. By utilizing real-time sensed data, ANGEL is capable to assess the system behavior and simulate complex attacks. As a result, defense and optimization strategies are retrieved for the physical system to form a self-healing grid. By combining with edge computing technology, Maheswaran \textit{et al.} \cite{maheswaran2019fog} introduce a distributed DT network deployed at the edge machine where multiple AI models are hosted to solve different application problems, which enables the autonomous vehicle assistance system with low-latency and multiple applications. In the field of medical health, as a future development direction, \cite{6Gwireless} also mentions that as a further development of current medical simulation such as monitoring of macro-physical indicators and the prevention of dominant diseases, a complete DT based human body can be realized to simulate medication and to monitor the body reaction, which is a powerful tool to promote drug research, epidemic treatment and disease prediction, \emph{etc}.

\subsection{Computing and Analysis}

As a light-weight application scenario, computing and analysis of DT mainly focus on specific task implementation. It is more flexible and is often used in conjunction with simulation scenarios by designing corresponding AI models using techniques such as deep learning and transfer learning to simulate real-world target tasks. For example, by deploying a retrained YOLO model, a DT system can be used to detect specified targets, such as pedestrian, cars, indicators and special equipment. Typically, to maintain the privacy of DT system, computing and analysis applications are required to only contain a bidirectional intra-twin communication link to transmit data and information with simulation platform, while the simulation platform has the permission to determine whether to exchange data with external devices according to the task requirements.

With the proliferation of AI technologies, multiple AI based methods are proposed to improve the performance of DT system. By utilizing deep learning and computer aided engineering simulation, \cite{FRANCIOSA2020369} develops an equipment parts assembly model hosted in DT system, which effectively improves the accuracy of equipment welding. In the remote health monitoring area, \cite{10.1145/3410530.3414316} introduces multiple machine learning models to predict and evaluate individual stress level according to washable sensors; in this way, the stress information can be learned and analyzed without exposing humans to negative environment. To fill the gap in the integration of DT based models with the observed cyber-physical system, \cite{10.1145/3365438.3410941} proposes a model-driven method that can fully analyze and describe the whole ecosystem of cyber-physical system and its DT, which helps eliminate repetitive programs and foster the systematic engineering of DT.

\section{Justification of Concepts}

This section compares DT with existing networking technologies.

\subsection{Comparison to Cloud Computing and Edge Computing}

The cloud computing is an integrated and open platform to host the data and
computing in the worldwide. Built upon a cluster of powerful computing
servers using the virtualization technology, the cloud computing provides a
global container over the world for users to deploy their applications and
services.

The edge computing is an extension of cloud computing which deploys
lightweight cloud-like devices at the user's premises so as to avoid the
remote long-thin connections from mobile users to the cloud, and provide the
fast-rate real-time responses to local service requests. Similar to cloud
computing, the edge computing provides an open container facility but to
local wireless devices only, where the wireless devices can offload their
applications and computing tasks.

In contrast to a hardware computing facility featured by the caching and computing
resource, the DT is a computing framework where the virtual part, \emph{i.e.}%
, DR, is a software application installed and operated on the cloud or edge serve.
Different from the cloud and edge computing which are open to different
users, the DR corresponds one-to-one to its physical part. As such, the DR
should have private data and protected storage and computing space in the
virtual space. To obtain a clearer understanding, we summarize the comparisons in Table~\ref{tab:1}.

\begin{table}[tbp]
\caption{Comparison of cloud, edge and DR}\centering%
\begin{tabular}{|c|c|c|c|}
\cline{1-4}
& Cloud & Edge & DR \\ \hline\hline
Hardware / Software & hardware & hardware & software \\ \cline{1-4}
Amount of Users & \emph{n} & \emph{n} & 1 \\ \cline{1-4}
Personalization & low & median & high \\ \cline{1-4}
Privacy & median & median & high \\ \cline{1-4}
Deployment & global & local & global/local/PE \\ \cline{1-4}
Data Carrying Capacity & high & median & low \\ \cline{1-4}
\end{tabular}%
\label{tab:1}
\end{table}

The DT system provides a secured framework for communications. In a complete
DT communication system, each PE has its own exclusive DR, and the data
communication between them follows the private principle, which can be used
to ensure sufficient security and privacy. Specifically, in intra-twin
communication, DR and PE first establishes a private link in the process of
data transmission and sharing, and all data only can be transmitted through
this specific link. During the inter-twin communication, all data first
request for the PE's permission. In this process, all private information in
PE will be retained, and only data marked as approved information can be
exchanged between DRs via inter-twin communications.

\begin{figure*}[tbp]
\centering
\includegraphics[width=\linewidth]{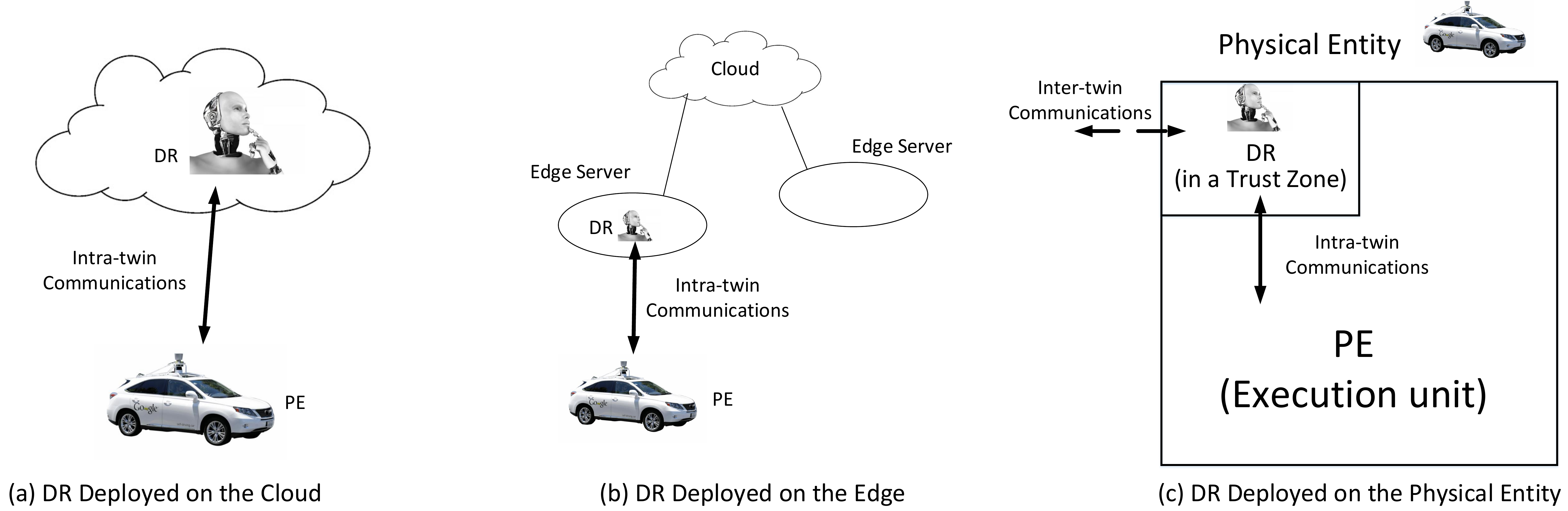}
\caption{Deployment of DR}
\label{fig: deployment}
\end{figure*}

\subsection{Comparison to Mobile Cloud Computing}

The mobile cloud computing is the most similar framework to DT. Mobile cloud
computing also maintains a private application on the cloud dedicated to
mobile users, \emph{e.g.}, Apple iCloud. The design goal of mobile cloud
computing is to alleviate the computing and storage burden of mobile devices
with the always-on cloud resources.

The existing literature on mobile cloud computing mainly considers the
applications of data offloading by uploading and storing data from mobile
users to cloud to reduce the local load and cost \cite{6365155}. DT system
can be seen as an evolutionary version of mobile cloud. In addition to
transmitting and storing data, the DT system maintains an autonomous
software application on cloud in which DR can perform full AI applications,
such as data cleaning, data fusion, as well as data processing and analysis,
so as to analyze and predict the real scenarios faced by PE, and assist PE
towards the optimal decision making. In addition, compared with mobile cloud
computing, which works in a reactive manner, the DR can work in a proactive
manner in communications. For example, depending on the needs of different
tasks and deployment environments, DR can predict the communication
bandwidth of the intra-twin communications and actively schedule the data
synchronizations towards the optimal communication performance.

\subsection{Comparison to Artificial Intelligence}

Artificial intelligence refers to the technology that simulates human
intelligence through computer programs. By leveraging the mathematical
optimization and logical reasoning, AI can learn external data and
adaptively generate solutions to achieve specific goals, such as planning,
perceiving and manipulating machinery. Different algorithmic models can be
integrated to suit the needs of tasks in complex scenarios. In a nutshell,
AI is a task-driven technique. It treats models as the basic unit for
achieving functions and only solves one problem at a time.

DT is an integrated multiphysics that looks at the big picture and provides
a holistic solution for large-scale tasks. A DR can contain multiple AI
models. According to the answers of sub-problems given by different models,
the DR can eventually provide a comprehensive output to optimize the
solution \cite{zhuangpeijie2018patent}. For example, in the process of
moving goods by a large robotic arm, an RNN based speech recognition model
can recognize the voice commands issued by the user, and then, through
semantic analysis, the target product, action command, and movement
destination information can be recognized and transmitted as input to the
corresponding target detection model, action planning model, and path
planning model. Finally, through the analysis and calculation in DR, a
strategy plan with the least time and material cost consumption can be
transmitted to the PE to achieve the target task.

\section{Deployment of DR}

As a light-weight software solution, DT communication system is actually quite
flexible; it can be deployed in multiple environment and through different
networks, as shown in Fig.~\ref{fig: deployment}. The descriptions and
examples are detailed as follows.

\subsection{Deployment on the Cloud}

As the primary way of the DT system deployment, in most cases, DRs are
deployed in a cloud-based environment. This is suitable for the physical
entities that need to constantly move or that need to be called by multiple
devices. For example, in the application of healthcare, most of the medical
monitor devices is required to be moved to different area based on the
demand. In addition, the wearable devices also should move at will with the
wearer. In this case, to ensure the stability of data transmission and
reduce jitter, DR can be deployed in the cloud to cope with the movement of
devices on the PE side and to ensure the interaction of data between various
different monitoring devices \cite{zhanglin2018patent}.

\subsection{Deployment on the Edge}

Other than deploying the DR system on the cloud, the DT system can also be
built on edge environment, which is more suitable for large equipment that
does not require frequent location changes. Take remote surgery robot as
example, as a large equipment with high requirements for device precision
and sterile environment, the surgical robot is usually fixed in a specific
location in the operating room and does not need to be moved frequently.
More importantly, during the procedure, the robot and the surgeon uses a
private link for data and command communication so as to ensure that the
commands from surgeon and the feedback from robot movements are real-time
and uninterrupted. By deploying the DT system at the edge, not only can we
take advantage of the localized edge environment to improve the data processing and
analysis performance of AI models, but also combine the advantages of fast-rate low-cost edge connections to ensure the privacy and synchronization of data transmission \cite%
{rasheed2019digital}.

\subsection{Deployment on the Physical Entity}

Other than the deployment approach introduced above, the DT system also can
be deployed directly on the reserved trust area at the physical entity.
Specifically, in the real-world applications, especially industrial systems,
the hardware of physical entity needs to be protected from malicious
communications. In this case, a DR can be deployed at the trust zone of the
physical entity. Any outside communications need to go through the DR using
the inter-twin communications, and the verified and secured data are
forwarded from DR to the execution unit of PE through the private intra-twin
communications. In this scenario, a DR may perform the functions such as
data cleaning, security verification, honeypot, \emph{etc}.

\section{Open Research Issues}

Although the concept of DT communication system is increasingly pervasive in
recent years, it is still in its infancy. From the perspective of
intra-twin and inter-twin communications, some open issues are discussed as
follows.

\subsection{Intra-twin Communications}

\emph{Communication Efficiency}: The intra-twin communications are featured by the private, synchronized and wireless nature. As the PE would be moving constantly and dynamically connect to the network with unreliable and heterogenous wireless access technologies, to guarantee the reliable and timely data synchronization within the DT is challenging. Note that the intra-twin communications are long-lasting connections, it is possible for DR to learn and predict the access bandwidth of PE and adapt the data synchronization accordingly.

\emph{Security}: For the challenges of security requirements, the intra-twin communication
mainly focuses on the privacy protection and attack defense due to the
private data involved. These challenges can be solved through multiple
technologies such as blockchain, cryptography and AI. By collecting and
training attack data and abnormal behaviors, the system can be trained to
identify and even predict attack before it happens.

\subsection{Inter-twin Communications}

\emph{Communication Efficiency}: The inter-twin communications are typically through the Internet, or even shared memory when the DRs are in the same cloud. The communication bandwidth is therefore not the bottlecheck issue. Instead, how to enable DRs the locate the useful information source for data fetching is challenging for inter-twin communications. The named data networking is a plausible approach for DRs to locate the needed information.

\emph{Security}: In contrast to the security challenges of intra-twin communications,
inter-twin communication should be more concerned with the boundaries and
integrity of data transmission. That is, before transmitting information to
another DR, the DT system should first justify if the target data has
sufficient permissions to be disclosed. In addition, to ensure the integrity
of the data, when historical data is lost or incomplete, the blockchain can
be used on the DR of a security-related DT system to provide important input.

\section{Conclusion}

As a virtual digital equivalent of a physical system, DT communication
system presents many possibilities for innovation, such as give real-time
prediction and provide decision support for live operations. More
importantly, as an integrated AI solution, the application of DT system shows
better performance in mobile agent system.

In this paper, we describe the basic structure and advantages of the DT
system. By comparing it with cloud computing, mobile cloud computing and
edge computing, we explore the application scope of DT and discuss the
deployment approach in detail. Despite the research of DT is still in its
initial stage, we provide a new perspective for understanding DT and lay the
theoretical foundation for its in-depth use in mobile devices.

\bibliographystyle{IEEEtran}
\bibliography{ref}

\end{document}